\documentclass[conference]{IEEEtran}
\usepackage{amssymb, amsmath}
\usepackage{amsfonts}
\usepackage{bbm}
\usepackage{mathrsfs}
\usepackage{xspace}
\usepackage{bm}

\usepackage{cite}
\usepackage{epsfig}
\usepackage{cases}
\usepackage{psfrag}
\usepackage{enumerate}

\usepackage{makeidx}
\usepackage{supertabular}
\usepackage{acronym}

\newtheorem{lemma}{Lemma}

\newenvironment{textbmatrix}{   \setlength{\arraycolsep}{2.5pt}%
                                                                \big[\begin{matrix}}{\end{matrix}\big]%
                                                                \raisebox{0.08ex}{\vphantom{M}}}

\def\be{\begin{equation}}
\def\ee{\end{equation}}
\def\een{\nonumber \end{equation}}
\def\mat{\begin{bmatrix}}
\def\emat{\end{bmatrix}}
\def\btm{\begin{textbmatrix}}
\def\etm{\end{textbmatrix}}

\def\ba#1\ea{\begin{align}#1\end{align}}
\def\bs#1\es{\begin{split}#1\end{split}}
\def\bg#1\eg{\begin{gather}#1\end{gather}}
\def\bi#1\ei{\begin{itemize}#1\end{itemize}}

\newcommand{\safemath}[2]{\newcommand{#1}{\ensuremath{#2}\xspace}}

\DeclareMathOperator*{\argmin}{arg\;min}                
\DeclareMathOperator{\Exop}{\mathbb{E}}         

\safemath{\interior}{\mathrm{Int}}                       

\safemath{\dfn}{:=}                                                     
\safemath{\dirac}{\delta}                                       

\safemath{\No}{N_0}                                                     
\safemath{\Es}{E_s}                                                     
\safemath{\Eb}{E_b}                                                     
\safemath{\EbNo}{\frac{\Eb}{\No}}
\safemath{\EsNo}{\frac{\Es}{\No}}

\DeclareMathOperator{\CHop}{\ensuremath{\mathbb{H}}} 
\safemath{\tvir}{h_{\CHop}}                                     
\safemath{\tvtf}{L_{\CHop}}                                     
\safemath{\spf}{S_{\CHop}}
\safemath{\bff}{H_{\CHop}}                                      

\safemath{\ircf}{R_{h}}                                         
\safemath{\scf}{R_{S}}                                          
\safemath{\tfcf}{R_{L}}                                         
\safemath{\bfcf}{R_{H}}                                         

\safemath{\mi}{I}                                                       
\safemath{\capacity}{C}                                         

\newcommand{\pdf[2]}{f_{{#1}}\left({#2}\right)}                    

\safemath{\uniform}{\mathcal{U}}                        
\safemath{\normal}{\mathcal{N}}                         
\safemath{\circnorm}{\mathcal{CN}}                      
\safemath{\mchain}{\leftrightarrow}                     

\safemath{\dB}{\,\mathrm{dB}}
\safemath{\dBm}{\,\mathrm{dBm}}
\safemath{\Hz}{\,\mathrm{Hz}}
\safemath{\kHz}{\,\mathrm{kHz}}
\safemath{\MHz}{\,\mathrm{MHz}}
\safemath{\GHz}{\,\mathrm{GHz}}
\safemath{\s}{\,\mathrm{s}}
\safemath{\ms}{\,\mathrm{ms}}
\safemath{\mus}{\,\mathrm{\mu s}}
\safemath{\ns}{\,\mathrm{ns}}
\safemath{\meter}{\,\mathrm{m}}
\safemath{\mm}{\,\mathrm{mm}}
\safemath{\cm}{\,\mathrm{cm}}
\safemath{\m}{\,\mathrm{m}}
\safemath{\W}{\,\mathrm{W}}
\safemath{\J}{\,\mathrm{J}}
\safemath{\K}{\,\mathrm{K}}
\safemath{\bit}{\,\mathrm{bit}}
\safemath{\mW}{\,\mathrm{mW}}
\safemath{\nW}{\,\mathrm{nW}}
\safemath{\pW}{\,\mathrm{pW}}
\safemath{\muW}{\,\mu\mathrm{W}}
\safemath{\Watt}{\,\mathrm{W}}
\safemath{\kbps}{\,\mathrm{kb/s}}
\safemath{\Mbps}{\,\mathrm{Mb/s}}

\safemath{\define}{\triangleq}                  

\safemath{\equivalent}{\sim}
\safemath{\distas}{\sim}                                        

\safemath{\reals}{\mathbb{R}}
\safemath{\positivereals}{\mathbb{R}^{+}}
\safemath{\integers}{\mathbb{Z}}
\safemath{\posint}{\mathbb{Z}_{+}}
\safemath{\naturals}{\mathbb{N}}
\safemath{\complexset}{\mathbb{C}}

\safemath{\setA}{\mathcal{A}}
\safemath{\setB}{\mathcal{B}}
\safemath{\setC}{\mathcal{C}}
\safemath{\setD}{\mathcal{D}}
\safemath{\setE}{\mathcal{E}}
\safemath{\setF}{\mathcal{F}}
\safemath{\setG}{\mathcal{G}}
\safemath{\setH}{\mathcal{H}}
\safemath{\setI}{\mathcal{I}}
\safemath{\setJ}{\mathcal{J}}
\safemath{\setK}{\mathcal{K}}
\safemath{\setL}{\mathcal{L}}
\safemath{\setM}{\mathcal{M}}
\safemath{\setN}{\mathcal{N}}
\safemath{\setO}{\mathcal{O}}
\safemath{\setP}{\mathcal{P}}
\safemath{\setQ}{\mathcal{Q}}
\safemath{\setR}{\mathcal{R}}
\safemath{\setS}{\mathcal{S}}
\safemath{\setT}{\mathcal{T}}
\safemath{\setU}{\mathcal{U}}
\safemath{\setV}{\mathcal{V}}
\safemath{\setW}{\mathcal{W}}
\safemath{\setX}{\mathcal{X}}
\safemath{\setY}{\mathcal{Y}}
\safemath{\setZ}{\mathcal{Z}}
\safemath{\emptySet}{\varnothing}

\safemath{\bma}{\mathbf{a}}
\safemath{\bmb}{\mathbf{b}}
\safemath{\bmc}{\mathbf{c}}
\safemath{\bmd}{\mathbf{d}}
\safemath{\bme}{\mathbf{e}}
\safemath{\bmf}{\mathbf{f}}
\safemath{\bmg}{\mathbf{g}}
\safemath{\bmh}{\mathbf{h}}
\safemath{\bmi}{\mathbf{i}}
\safemath{\bmj}{\mathbf{j}}
\safemath{\bmk}{\mathbf{k}}
\safemath{\bml}{\mathbf{l}}
\safemath{\bmm}{\mathbf{m}}
\safemath{\bmn}{\mathbf{n}}
\safemath{\bmo}{\mathbf{o}}
\safemath{\bmp}{\mathbf{p}}
\safemath{\bmq}{\mathbf{q}}
\safemath{\bmr}{\mathbf{r}}
\safemath{\bms}{\mathbf{s}}
\safemath{\bmt}{\mathbf{t}}
\safemath{\bmu}{\mathbf{u}}
\safemath{\bmv}{\mathbf{v}}
\safemath{\bmw}{\mathbf{w}}
\safemath{\bmx}{\mathbf{x}}
\safemath{\bmy}{\mathbf{y}}
\safemath{\bmz}{\mathbf{z}}

\bmdefine{\biad}{a}
\bmdefine{\bibd}{b}
\bmdefine{\bicd}{c}
\bmdefine{\bidd}{d}
\bmdefine{\bied}{e}
\bmdefine{\bifd}{f}
\bmdefine{\bigd}{g}
\bmdefine{\bihd}{h}
\bmdefine{\biid}{i}
\bmdefine{\bijd}{j}
\bmdefine{\bikd}{k}
\bmdefine{\bild}{l}
\bmdefine{\bimd}{m}
\bmdefine{\bind}{n}
\bmdefine{\biod}{o}
\bmdefine{\bipd}{p}
\bmdefine{\biqd}{q}
\bmdefine{\bird}{r}
\bmdefine{\bisd}{s}
\bmdefine{\bitd}{t}
\bmdefine{\biud}{u}
\bmdefine{\bivd}{v}
\bmdefine{\biwd}{w}
\bmdefine{\bixd}{x}
\bmdefine{\biyd}{y}
\bmdefine{\bizd}{z}

\bmdefine{\bixid}{\xi}
\bmdefine{\bilambdad}{\lambda}
\bmdefine{\bimud}{\mu}
\bmdefine{\bithetad}{\theta}
\bmdefine{\biphid}{\phi}

\safemath{\bmia}{\biad}
\safemath{\bmib}{\bibd}
\safemath{\bmic}{\bicd}
\safemath{\bmid}{\bidd}
\safemath{\bmie}{\bied}
\safemath{\bmif}{\bifd}
\safemath{\bmig}{\bigd}
\safemath{\bmih}{\bihd}
\safemath{\bmii}{\biid}
\safemath{\bmij}{\bijd}
\safemath{\bmik}{\bikd}
\safemath{\bmil}{\bild}
\safemath{\bmim}{\bimd}
\safemath{\bmin}{\bind}
\safemath{\bmio}{\biod}
\safemath{\bmip}{\bipd}
\safemath{\bmiq}{\biqd}
\safemath{\bmir}{\bird}
\safemath{\bmis}{\bisd}
\safemath{\bmit}{\bitd}
\safemath{\bmiu}{\biud}
\safemath{\bmiv}{\bivd}
\safemath{\bmiw}{\biwd}
\safemath{\bmix}{\bixd}
\safemath{\bmiy}{\biyd}
\safemath{\bmiz}{\bizd}

\safemath{\bmxi}{\bixid}
\safemath{\bmlambda}{\bilambdad}
\safemath{\bmmu}{\bimud}
\safemath{\bmtheta}{\bithetad}
\safemath{\bmphi}{\biphid}

\safemath{\bA}{\mathbf{A}}
\safemath{\bB}{\mathbf{B}}
\safemath{\bC}{\mathbf{C}}
\safemath{\bD}{\mathbf{D}}
\safemath{\bE}{\mathbf{E}}
\safemath{\bF}{\mathbf{F}}
\safemath{\bG}{\mathbf{G}}
\safemath{\bH}{\mathbf{H}}
\safemath{\bI}{\mathbf{I}}
\safemath{\bJ}{\mathbf{J}}
\safemath{\bK}{\mathbf{K}}
\safemath{\bL}{\mathbf{L}}
\safemath{\bM}{\mathbf{M}}
\safemath{\bN}{\mathbf{N}}
\safemath{\bO}{\mathbf{O}}
\safemath{\bP}{\mathbf{P}}
\safemath{\bQ}{\mathbf{Q}}
\safemath{\bR}{\mathbf{R}}
\safemath{\bS}{\mathbf{S}}
\safemath{\bT}{\mathbf{T}}
\safemath{\bU}{\mathbf{U}}
\safemath{\bV}{\mathbf{V}}
\safemath{\bW}{\mathbf{W}}
\safemath{\bX}{\mathbf{X}}
\safemath{\bY}{\mathbf{Y}}
\safemath{\bZ}{\mathbf{Z}}

\bmdefine{\biAd}{A}
\bmdefine{\biBd}{B}
\bmdefine{\biCd}{C}
\bmdefine{\biDd}{D}
\bmdefine{\biEd}{E}
\bmdefine{\biFd}{F}
\bmdefine{\biGd}{G}
\bmdefine{\biHd}{H}
\bmdefine{\biId}{I}
\bmdefine{\biJd}{J}
\bmdefine{\biKd}{K}
\bmdefine{\biLd}{L}
\bmdefine{\biMd}{M}
\bmdefine{\biOd}{N}
\bmdefine{\biPd}{O}
\bmdefine{\biQd}{P}
\bmdefine{\biRd}{R}
\bmdefine{\biSd}{S}
\bmdefine{\biTd}{T}
\bmdefine{\biUd}{U}
\bmdefine{\biVd}{V}
\bmdefine{\biWd}{W}
\bmdefine{\biXd}{X}
\bmdefine{\biYd}{Y}
\bmdefine{\biZd}{Z}

\bmdefine{\biDelta}{\Delta}
\bmdefine{\biLambda}{\Lambda}
\bmdefine{\biPhi}{\Phi}
\bmdefine{\biSigma}{\Sigma}
\bmdefine{\biOmega}{\Omega}
\bmdefine{\biTheta}{\Theta}

\safemath{\bimA}{\biAd}
\safemath{\bimB}{\biBd}
\safemath{\bimC}{\biCd}
\safemath{\bimD}{\biDd}
\safemath{\bimE}{\biEd}
\safemath{\bimF}{\biFd}
\safemath{\bimG}{\biGd}
\safemath{\bimH}{\biHd}
\safemath{\bimI}{\biId}
\safemath{\bimJ}{\biJd}
\safemath{\bimK}{\biKd}
\safemath{\bimL}{\biLd}
\safemath{\bimM}{\biMd}
\safemath{\bimN}{\biNd}
\safemath{\bimO}{\biOd}
\safemath{\bimP}{\biPd}
\safemath{\bimQ}{\biQd}
\safemath{\bimR}{\biRd}
\safemath{\bimS}{\biSd}
\safemath{\bimT}{\biTd}
\safemath{\bimU}{\biUd}
\safemath{\bimV}{\biVd}
\safemath{\bimW}{\biWd}
\safemath{\bimX}{\biXd}
\safemath{\bimY}{\biYd}
\safemath{\bimZ}{\biZd}

\safemath{\bDelta}{\bielta}
\safemath{\bLambda}{\biLambda}
\safemath{\bPhi}{\biPhi}
\safemath{\bSigma}{\biSigma}
\safemath{\bOmega}{\biOmega}
\safemath{\bTheta}{\biTheta}

\safemath{\veca}{\bma}
\safemath{\vecb}{\bmb}
\safemath{\vecc}{\bmc}
\safemath{\vecd}{\bmd}
\safemath{\vece}{\bme}
\safemath{\vecf}{\bmf}
\safemath{\vecg}{\bmg}
\safemath{\vech}{\bmh}
\safemath{\veci}{\bmi}
\safemath{\vecj}{\bmj}
\safemath{\veck}{\bmk}
\safemath{\vecl}{\bml}
\safemath{\vecm}{\bmm}
\safemath{\vecn}{\bmn}
\safemath{\veco}{\bmo}
\safemath{\vecp}{\bmp}
\safemath{\vecq}{\bmq}
\safemath{\vecr}{\bmr}
\safemath{\vecs}{\bms}
\safemath{\vect}{\bmt}
\safemath{\vecu}{\bmu}
\safemath{\vecv}{\bmv}
\safemath{\vecw}{\bmw}
\safemath{\vecx}{\bmx}
\safemath{\vecy}{\bmy}
\safemath{\vecz}{\bmz}
\safemath{\vecZero}{\bZero}

\safemath{\vecxi}{\bmxi}
\safemath{\veclambda}{\bmlambda}
\safemath{\vecmu}{\bmmu}
\safemath{\vectheta}{\bmtheta}
\safemath{\vecphi}{\bmphi}

\safemath{\matA}{\bA}
\safemath{\matB}{\bB}
\safemath{\matC}{\bC}
\safemath{\matD}{\bD}
\safemath{\matE}{\bE}
\safemath{\matF}{\bF}
\safemath{\matG}{\bG}
\safemath{\matH}{\bH}
\safemath{\matI}{\bI}
\safemath{\matJ}{\bJ}
\safemath{\matK}{\bK}
\safemath{\matL}{\bL}
\safemath{\matM}{\bM}
\safemath{\matN}{\bN}
\safemath{\matO}{\bO}
\safemath{\matP}{\bP}
\safemath{\matQ}{\bQ}
\safemath{\matR}{\bR}
\safemath{\matS}{\bS}
\safemath{\matT}{\bT}
\safemath{\matU}{\bU}
\safemath{\matV}{\bV}
\safemath{\matW}{\bW}
\safemath{\matX}{\bX}
\safemath{\matY}{\bY}
\safemath{\matZ}{\bZ}
\safemath{\matZero}{\bZero}

\safemath{\matDelta}{\bDelta}
\safemath{\matLambda}{\bLambda}
\safemath{\matPhi}{\bPhi}
\safemath{\matSigma}{\bSigma}
\safemath{\matOmega}{\bOmega}
\safemath{\matTheta}{\bTheta}

\safemath{\matIdentity}{\matI}

\renewcommand{\figurename}{Fig.}

\renewcommand{\tablename}{Table}

\usepackage{fancyhdr, color, listings, amsbsy, fixmath, amssymb}
\usepackage[usenames,dvipsnames]{xcolor}
\usepackage{float}
\usepackage{mathtools}
\usepackage{amsmath}
\usepackage{dsfont}
\usepackage{amsbsy}

\usepackage{graphicx}
\usepackage{psfrag}
\usepackage[tight,footnotesize]{subfigure}
\usepackage{empheq}
\usepackage{xurl}

\safemath{\bps}{\mathrm{bits/s}}
\safemath{\bJoule}{\mathrm{bits/J}}
\safemath{\MbJoule}{\mathrm{Mbits/J}}
\safemath{\KbJoule}{\mathrm{Kbits/J}}
\safemath{\GbJoule}{\mathrm{Gbits/J}}
\safemath{\Gb}{\mathrm{Gbits}}
\safemath{\Mb}{\mathrm{Mbits/s}}
\safemath{\Joule}{\mathrm{Joule}}
\safemath{\Kg}{\mathrm{Kg}}
\safemath{\mps}{\mathrm{m/s^2}\!}

\usepackage[linesnumbered,lined,boxed, ruled]{algorithm2e}

\newcommand{\Set}{\textbf{set }}
\newcommand{\Update}{\textbf{update }}
\newcommand{\Compute}{\textbf{compute }}

\newcommand{\Solve}{\textbf{solve }}
\newcommand{\Init}{\textbf{initialize }}

\IEEEoverridecommandlockouts

\begin{document}

\title{Joint Latency-Energy Minimization for Fog-Assisted Wireless IoT Networks}

\author{Farshad Shams$^{*1}$, Vincenzo Lottici$^1$, Zhi Tian$^2$, and Filippo Giannetti$^1$
\thanks{$^*$Corresonding author.}
\thanks{$^1$Dept. of Information Engineering, University of Pisa, via Caruso 16, 56126 Pisa, Italy. (email: farshad.shams@ing.unipi.it, \{vincenzo.lottici, filippo.giannetti\}@unipi.it)}
\thanks{$^2$Dept. of Electrical and Computer Engineering, George Mason University, Fairfax, VA, 22030, USA.  (email: ztian1@gmu.edu)}
}


\maketitle

\begin{abstract}
This work aims to present a joint resource allocation method for a fog-assisted network wherein IoT wireless devices simultaneously offload their tasks to a serving fog node.
The main contribution is to formulate joint minimization  of service latency and energy consumption objectives subject to both radio and computing constraints.
Moreover, unlike previous works that set a fixed value to the circuit power dissipated to operate a wireless device, practical  models are considered.
To derive the Pareto boundary  between two conflicting objectives we consider, Tchebyshev theorem is used for each wireless device.
The competition among devices is modeled using the cooperative Nash bargaining solution and its unique cooperative Nash equilibrium (NE) is computed  based on block coordinate descent algorithm.
Numerical results obtained using realistic models are presented to corroborate the effectiveness of the proposed algorithm.
\end{abstract}


\section{Introduction}

With the day-to-day increasing popularity of wireless IoT devices, the number of connections, and consequently, the progress of mobile computing is rapidly rising.
Accordingly, the issues of response time and energy consumption come out as very critical ones.
In order to address and suggest competitive solutions,
a \emph{fog-assisted network} has been proposed to manage resource-demanding delay-sensitive next-generation services,
such as those required by (resource-limited) wireless devices, e.g.,
 mobile phones, surveillance video cameras, or sensors for measuring physical variables (pressure, temperature, humidity, and so on).
According to this solution, the \emph{fog nodes}
are located as close as possible to where data are produced to enable more efficient computing, storage, and networking functions for  resource-limited IoT devices \cite{bonomi12}.
Hence, the offloading of resource-consuming and delay-sensitive tasks from end IoT devices to a nearby fog computing node can be efficiently performed.
In this work, a single-cell fog-assisted network is considered where the base-station (BS) is equipped with one fog node to  serve a number of IoT devices.
In other words, IoT devices simultaneously communicate with the BS to offload their tasks on the fog node which executes them by sharing its computing resources.
The objective of interest is to develop an efficient resource allocation algorithm for joint energy consumption and service latency minimization of the IoT devices.

\subsection{Motivation}
As the number of resource-constrained IoT devices continues to grow, there is an increasing need for
efficient computing and communications in wireless networks \cite{hossain12}.
Task offloading to fog nodes has emerged as a promising solution for enhancing the computing capabilities
of IoT devices distributed across a network.
However, fog-assisted networks, serving as infrastructure for both communication and computing services,
need to enhance the efficiency of both physical and application layers \cite{han08}.
Jointly optimizing the allocation of resources in both the physical and application layers of the network,
including transmit power levels and computing resources, can improve network performance in terms of both
objectives of interest, i.e., service latency and energy consumption.
Pursuing the above approach, we jointly minimize both objectives, thus making fog-assisted networks
an appealing solution  to support modern applications.
In this joint optimization, it is critical to derive the Pareto boundary:
it provides a tradeoff between the two conflicting objectives, thus allowing designers to
optimize the network configuration under the various constraints at both physical and application layers.
In nutshell, deriving the latency-energy Pareto boundary is a crucial step towards achieving efficient resource allocation in
fog-assisted IoT wireless networks.

\subsection{Related works}\label{sec:Related works}
In fog-assisted networks, a resource allocation algorithm focuses on how to optimally allocate resources of physical and application layers to resource-constrained wireless IoT devices.
Most resource allocation research works address the minimization of the end-to-end delay or energy consumption of the task offloading and computing.
In the literature of resource allocation, in general, one can find three types of objective function optimizations:
($i$) minimizing the overall energy required for task offloading and execution while satisfying a service latency constraint,
($ii$) minimizing the service latency subject to a constraint on available energy,
and ($iii$) minimizing a weighted-sum of the energy consumption and the service latency.
As fog-assisted networks are designed for both communication and computing services with limited resources, the objective function of type ($iii$) seems to be more attractive in the literature. However, the demanding issue is to find the best weights such that the service latency and energy consumption of all devices is  jointly minimized.

For the objective function of type ($i$),  \cite{jeong18} deals with multi-cell single-antenna wireless device communication networks, while
\cite{alshuwaili17} investigates the problem in multi-cell MIMO networks.
For single-cell MIMO communications, \cite{mao16} and \cite{lagen18}  derive a (quasi) closed-form solution.
The mentioned works use centralized convex optimization techniques, based on successive convex approximation (SCA) \cite{scutari17} to minimize the sum of objective functions of all wireless devices (sum-utility).
Works \cite{marbukh19, lee19, chen19} formulate the same problem for a single-cell network consisting of mobile devices. In order to handle this matter, \cite{marbukh19} proposes a distributed greedy optimization algorithm, while \cite{lee19, chen19} propose a learning based algorithm to guarantee the service latency requirement of mobile devices.

For the objective function of type ($iii$), \cite{lyu18} and \cite{tran19} investigate the problem in a single-antenna network system.
Sub-optimal  solutions  are developed  by  decoupling  offloading  decisions  and  resource allocations. Again, all mentioned research works use SCA technique to minimize the sum of objective functions of all wireless devices.

There exist also some other works for multi-tiered systems where each task can be executed either on a nearby fog node or on an upper-tier node with greater computational capacity, e.g., on a cloud server.
Works \cite{du18, mukherjee19, schichao19, hung21} tackle the task offloading decision problem with the objective function of type ($ii$) by only  adjusting the transmit power levels. In \cite{plachy21} the same problem is formulated for a multi-cell network consisting of one mobile device.
In \cite{gao20} and \cite{tong20}, an objective function of type ($i$) is defined to minimize the overall energy consumption. They formulate the problem as a stochastic network optimization and a mix-integer non-linear programming, respectively.


The competition among wireless devices to access radio and computing resources can be modeled using game theory \cite{osborne94, han12}.
Using non-cooperative games, \cite{zheng18, josilo17, li18, liu19, jie19, li19, birhanie20, bandyopadhyay20} consider the problem where each wireless device wishes to minimize its energy consumption.
However, in \cite{josilo17} the transmit powers of devices are given as constants and the proposed algorithm allocate only the computing resource on a fog node to IoT devices. Works \cite{liu19, jie19, birhanie20} use Stackelberg game to investigate the problem. To compute the NE, \cite{liu19} uses the Bellman equations while \cite{jie19} proposes a greedy round-robin algorithm. In \cite{bandyopadhyay20} and \cite{bahreini21}, an auction game approach is used.

The review of the existing related works shows that the existing problem formulations suffer from the following three major drawbacks:
\begin{enumerate}
	\item  Existing works did not consider a practical/industrial model for power consumption of electronic circuits of wireless devices. They usually set a fixed amount for it, while the power consumption of electronic circuits involved in communication, e.g., radio frequency (RF) and baseband (BB) electronic circuits, are functions of the transmit power level. Indeed, by considering practical models spectral and energy efficiencies change significantly when non-practical models are assumed \cite{chen11}.
\item Many existing works in fog-assisted networks have focused on optimizing a weighted-sum of energy consumption and service latency.
    However, this does not result in optimal joint minimization since the latter are dependent functions of both radio and computing resource variables  \cite{boyd04}.
    Furthermore, $i$)  the results obtained from a weighted-sum function can vary greatly depending on the chosen weight, and
    $ii$) the lack of a common measurement unit of the objective function  makes it unsuitable from a network performance perspective.
    Therefore, it is mandatory to formulate alternative objective functions that enable joint minimization of both service latency and energy consumption.
    On the other side, for wireless network designers, the Pareto boundary represents a very useful tool,
    not well-investigated in the literature,
    that provides  the optimal tradeoff between service latency and energy consumption.
    To be specific, this helps on determining the minimum possible service latency while expending a given available energy or vice versa \cite{han08}.
\item Existing works propose a sum-utility objective function, i.e., they optimize the sum of objective functions of all wireless devices. This approach might lead to unfair resource allocation, that is, the devices with poor channel conditions are assigned with low amount of resources,  and vice versa for the ones in good operation conditions \cite{han12}.
\end{enumerate}

\subsection{Contributions}
This paper aims to develop an efficient communication and computing resource allocation method to overcome the above existing drawbacks.
 More precisely, the following specific contributions are made.
\begin{enumerate}
\item An optimized resource allocation method is developed for the uplink of a single-cell fog-assisted wireless network, where each IoT device offloads its task to the fog node deployed at the BS. This method jointly optimizes the IoT transmit power levels and the fog node CPU capacity assigned to each task, taking into account both service latency and energy consumption in the objective function.
\item  Practical models for the energy consumption of IoT devices are exploited in the optimization considering the power consumption of all electronic circuits involved in communication and computing.
\item     The bi-objective latency-energy minimization is addressed by deriving the Pareto boundary using the weighted Tchebyshev method \cite{Kais99}. This provides valuable insights into the tradeoff between the service latency and energy consumption and highlights their importance in network design and optimization.
    By adopting joint multi-objective optimization, network designers can achieve a more balanced and optimal configuration that meets the needs of different stakeholders, such as users, service providers, and the wireless network industry \cite{bjornson14}.
\item To allocate  network resources among competitive wireless devices in an efficient and fair manner, the problem is formulated using a cooperative game approach based on the Nash bargaining solution (NBS) \cite{osborne94},  that guarantees the uniqueness and optimal performance at proportional fairness.\footnote{In its original definition, a vector $\{u_k^\star\}_{k=1}^K$ is said to be proportionally fair if, for any other feasible vector $\{u_k\}_{k=1}^K$, the sum of proportional changes $(u_k - u_k^\star)/u_k^\star$ is nonpositive \cite{boche09}.}
    Each point on the latency-energy Pareto boundary curve is thus given by the output the NBS.
    Specifically, to compute the unique cooperative Nash equilibrium (NE), an algorithm based on block coordinate descent \cite{wright15} is properly developed.
\item    To corroborate the proposed joint resource allocation algorithm, realistic numerical results are presented to quantify the performance of network IoT devices. Specifically, it is shown that considering a practical power consumption model instead of non-practical ones significantly changes the behavior of the Pareto boundary.
\end{enumerate}

\subsection{Structure of the paper}
The rest of the paper is organized as follows.  The system model and the problem statement are presented in Sect. \ref{sec:SystemModel}.
Service latency and energy consumption are computed in Sect. \ref{sec:ProblemForm}.
To derive the latency-energy Pareto boundary for each wireless device, in Sect. \ref{sec:Joint_Opt} a bi-objective optimization problem is formulated followed by the convexification of the non-convex constraints in Sect. \ref{sec:convexification}.
Then, in Sect. \ref{Sec:Coop_Game}, the competition among devices is modeled using a NBS-based approach and the unique cooperative NE point is  computed.
Numerical results are provided in Sect. \ref{sec:NumRes}, and
finally, concluding remarks are given in Sect. \ref{sec:Concl}.

\section{Single-Cell Network Model}\label{sec:SystemModel}

\begin{figure}[t]
	\begin{center}
		\psfrag{KDevices}[c][c][0.9]{$K$ wireless devices}
		\psfrag{KSubChannels}[c][c][0.9]{$K$ channels}
		\psfrag{MEC}[c][c][0.9]{Fog node}
		\includegraphics[width=0.4\textwidth]{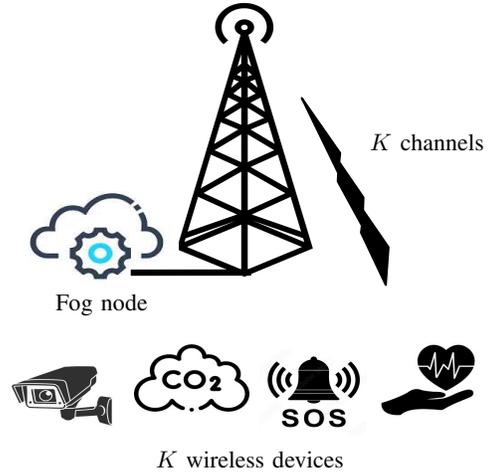}
	\end{center}
	\caption{The single-cell network model \cite{bonomi12}.}
	\label{fig:NetFig}
\end{figure}

As in \figurename~\ref{fig:NetFig} we address a single-cell network wherein one fog node is deployed on the BS to serve a number of  resource-constrained wireless devices which simultaneously offload their tasks.
The assumption of needing to offload all the tasks is realistic whenever the task execution requires, among others, data integration across multiple activities or interaction with a centralized database, e.g., a wide range of applications related to national security, health monitoring environment, and disaster management \cite{shams15}.

\subsection{Network parameters}
The number of $K$ wireless devices simultaneously communicate with the BS to offload their tasks on the fog node. The fog node executes all tasks by efficiently sharing its computing resources for different tasks.
The task of each device $k=1,\dots,\,K$ is defined by a 2-tuple $(D_k,\,C_k)$ where $D_k$ specifies the size of the task in [bits] and $C_k$ is the number of CPU cycles required for computing one-bit data in [CPU-cycle/bits] whose amount depends on specific computing-task complexities.
Assume that every task is atomic, i.e., it cannot be further broken down given strong dependence over different subtasks.

Devices are synchronized, i.e., they start to offload simultaneously at the fog node's request.
For task offloading, there are $K$ available wireless channels and devices are allocated with dedicated spectrums when they communicate with the BS, i.e., they do not interfere with each other, and the bandwidth is equal for all.
Each device $k$  adjusts its transmit power and then sends the whole task to the fog node.
The fog node assigns an individual virtual machine (VM) for each device, using VM multiplexing and consolidation techniques that allow multi-task parallel computation.
The fog node needs to adjust its CPU frequency for each VM, which executes the tasks and sends
the results to the device. Moreover, assuming relatively small sizes of computation results for applications (such as object detection in a video) and high transmission power at the BS, downloading is much faster than offloading and consumes negligible device energy.
Under these conditions, it is assumed that the downloading phase from the BS to wireless devices has a fixed duration.

Let us denote by $p_k$ in [W] the transmit power level of device $k$ that is fixed during task offloading. The transmit power level $p_k$ is upper-bounded by $\overline{p}_k$. The transmit power vector of all devices is denoted by $\mathbf{p} = \{p_k\}_{k=1}^K$.
It is also denoted by $\overline{f}_0$ the maximum computational capacity, in [CPU-cycles/s] (or [Hz]), of the fog node that will be shared by the devices.
For execution of a device $k$'s task, the fog node has to adjust the CPU frequency cycle of the assigned VM to $f_k$ subject to $\sum_{k=1}^K f_k \le \overline{f}_0$.  The CPU frequency adjustment vector is denoted by $\mathbf{f} = \{f_k\}_{k=1}^K$.
The goal is to jointly optimize power vector $\mathbf{p}$ and frequency vector $\mathbf{f}$ to jointly minimize both service latency and energy consumption of all devices.

\subsection{Service latency and energy consumption}
For a device $k$, the total service latency $T_k$ includes both: $1)$ the time $T_{k}^{\textrm{tx}}$ it needs to offload its task to the fog node and $2)$ The assigned VM takes execution time of $T_{k}^{\textrm{ex}}$ to execute the task.
The total energy consumption $E_k$ represents the total amount of radiative and non-radiative power consumption at the device. Unlike existing works, practical models are used for computing the power consumption of electronic circuits involved in the communication. The total energy consumption includes:
\begin{enumerate}
  \item $E_k^{\textrm{tx}}$: the total amount of power consumed over the duration of $T_k^{\textrm{tx}}$ to offload the task. It includes both radiative  transmit power level and the non-radiative power consumed by the RF and BB electronic circuits in the connected mode,
  \item $E_{k}^{\textrm{ex}}$: the energy consumed by the assigned VM on the fog node to execute the task,
  \item $E_k^{\textrm{on}}$: the power required to keep the device functioning within the whole service latency $T_k$. It includes all electronic circuits which keep a device on.
\end{enumerate}

\subsection{Latency-energy tradeoff}

To handle radio and computing resource allocations in wireless fog-assisted networks, the key performance objectives are service latency and energy consumption.
As it can be seen, both objectives depend on both physical layer and application layer resources; i.e., transmit power levels of the devices and CPU frequency assigned to VMs.
The goal is to optimally adjust $\mathbf{p}$ and $\mathbf{f}$ to execute all tasks in minimum time at the expense of minimum energy.
The trade-off is between service latency and energy consumption.
On the fog node, increasing the CPU frequency of each VM reduces the execution time. Meanwhile, it increases the energy consumption by the CPU.
Besides, increasing the transmit power level at a device increases the data-rate, and in turn, reduces the task offloading time. On the other hand, from an energy consumption point of view, it increases the communication cost.
As it can be perceived, service latency and energy consumption performances are correlated and are in essence conflicting.
Therefore, to achieve high computation and energy performance while respecting the imposed constraints and limitations, it is required to optimize radio and computation resource allocations jointly.
In this work, the goal is to jointly minimize the service latency and the energy consumption by adjusting the transmit power at each device and the CPU frequency of each VM.

\subsection{Wireless communication model }
As validated by the channel measurements reported in \cite{kermoal02}, if there is sufficient
scattering between a wireless device $k$ and the BS,  then the channel transfer function, in general, is modeled as $h_k = \sqrt{l_k}\,\epsilon_k$ where $\epsilon_k \in \mathcal{CN}(0,\,1)$ accounts for the
small-scale fading channel and $l_k$ describes the large-scale fading channel at distance $d_k$ (measured in [km]), computed as \cite{tse05}
\begin{equation}
	l_k = \frac{\beta}{\left(d_k\right)^\alpha}
\end{equation}
which includes distance-dependent path loss, shadowing, antenna gains, and penetration losses in propagation.
The parameter $\alpha$ is the path loss exponent and $\beta$ accounts for the overall path loss and shadowing effects from the device to the BS  at a reference distance of $1$ km. Since it is assumed devices are fixed, the amount of $l_k$ is fixed during the transmission.

To compute the data-rate during the task offloading, as time evolution is involved, the ergodic data-rate formula has to be used \cite{tse05} that is computed by
\begin{multline}
	\Exop\left\{ B \log_2\left(1 + \frac{p_k |h_k|^2}{BN_0}\right)\right\}  = \Exop\left\{ B \log_2\left(1 + \frac{p_k l_k|\epsilon_k|^2}{BN_0}\right)\right\}\\
	\le  B \log_2\left(1 + \frac{p_k l_k \Exop\{|\epsilon_k|^2\}}{BN_0}\right)  \qquad \text{[\bps]}
\end{multline}
where $B$ denotes the bandwidth assigned to each device in [Hz] and $N_0$ denotes the noise spectral density in [W/Hz]. The last inequality is a direct consequence of Jensen’s inequality \cite{boyd04}.
The duration of $T_k^{\textrm{tx}}$ is assumed large enough to set the variance value of the experienced fast fading terms to one.
In the above circumstances and for mathematical tractability, the following (upper-bound) formula is used for the data-rate of a device $k$
\begin{equation}\label{eq:sinr_k}
	R_k\left(p_k\right) = B\log_2\left(1 + \frac{p_k l_k}{BN_0}\right) \qquad \text{[\bps]}.
\end{equation}

\section{Service latency and energy consumption formulation}\label{sec:ProblemForm}
In the following, different terms of service latency and energy consumption are computed.

\subsection{Service latency}
For each device $k$, the total service latency $T_k$ includes both the time necessary to offload and execute the task.

{\boldmath\textbf{{Computing $T_k^{\textrm{tx}}$}}}:
For each device $k$, transmitting at data-rate $R_k(p_k)$ [\bps], the time necessary to transfer $D_k$ bits to the fog node is computed by
\begin{equation}\label{eq:T_k_tr}
T_k^{\textrm{tx}}\left(p_k\right) = \frac{D_k}{R_k\left(p_k\right)}\qquad[\s]
\end{equation}

{\boldmath\textbf{{Computing $T_k^{\textrm{ex}}$}}}:
Since operating at a constant CPU-cycle frequency is most energy-efficient, here, it is assumed the CPU cycle assigned to VMs remains fixed during task execution. The duration time of computing a task of $D_k$ bits is thus computed by
\begin{equation}\label{eq:T_k_exe}
T_k^{\textrm{ex}}\left(f_k\right) = \frac{C_k D_k}{f_k}\qquad[\s]
\end{equation}
Therefore, The total service latency is computed by
\begin{equation}\label{eq:T_k}
T_k\left(f_k,\,p_k\right) = T_k^{\textrm{tx}}\left(p_k\right) + T_k^{\textrm{ex}}\left(f_k\right)\; \quad[\s]
\end{equation}

\subsection{Energy consumption}
Total energy consumption includes the energy consumption for offloading a task, the energy consumption by computation resources, and the energy consumption by electronic circuits of the device.

{\boldmath\textbf{{Computing $E_k^{\textrm{tx}}$}}}: The total radiative and non-radiative energy consumption for task offloading is evaluated based on the device power consumption model proposed by \cite{jensen12, mads14, musiige12}. In the adopted model, the energy consumption for transmission includes transmit power level and network electronic circuits. It is evaluated as
\begin{equation}\label{eq:E_k_TXc}
E_k^{\textrm{tx}}\left(p_k\right) = \left(P_k^{c} + P_k^{BB} + P_k^{RF} + p_k\right)T_k^{\textrm{tx}}\left(p_k\right)\qquad[\Joule]
\end{equation}
where
\begin{itemize}
  \item $P_k^{c}$ is the power consumption of active transmission chain in the connected mode (e.g. power amplifier). Based on experimental assessments in \cite{jensen12}, it is equal to $1.35$ W.
  \item $P_k^{BB}$ is power consumption of the BB components (e.g. encoder circuit). It depends on the uplink data rate $R_k(p_k)$, in turns $p_k$, following the corresponding equation in Table \ref{tab:TX_E}.
      Note that, in the corresponding equation in Table \ref{tab:TX_E}, the value of input parameter $R_k(p_k)$ is in [Mbits/s] and the result of $P_k^{BB}$ is in [mW] that needs to be converted to [W] in \eqref{eq:E_k_TXc}.
  \item $P_k^{RF}$  is power consumption of radio RF components (e.g. modulation circuit). It depends on the transmission power $p_k$ following the equations in Table \ref{tab:TX_E} that is derived from \cite{musiige12}.
  \item $p_k$: is the radiative transmit power level during the time required for task offloading.
\end{itemize}

\begin{table}
  \centering
  \caption{Equations of the power consumptions $P_k^{BB}$ and $P_k^{RF}$ \cite{jensen12, musiige12}.}
  \vspace{-4mm}
\begin{tabular}{|c|c|c|c|}
\hline
  Part  & Variable & Model   \\
    \hline\hline
  $P_k^{BB}$ in [mW]& $R_k\left(p_k\right)$ in [Mbit/s] & $2110 + 0.87R_k\left(p_k\right)$  \\\hline
  $P_k^{RF}$ in [W]& $p_k$ in [W] & $ 0.6 + 10.1p_k$   \\
  \hline
\end{tabular}
  \label{tab:TX_E}
\end{table}

{\boldmath\textbf{{Computing $E_{k}^{\textrm{ex}}$}}}:
 Under the assumption of low CPU voltage, the energy consumption for each CPU cycle at frequency $f$ is computed by $E^{cyc}(f) = \lambda f^2$ \cite{burd96} in [Joule/CPU-cycle], where $\lambda$ is a constant determined by the computation electronic circuits. Therefore, energy consumption to execute a device's task on its VM is computed by
\begin{equation}\label{eq:E_k_exe}
E_k^{\textrm{ex}}\left(f_k\right) = \lambda f_k^2 C_k D_k \qquad [\Joule]
\end{equation}

{\boldmath\textbf{{Computing $E_k^{\textrm{on}}$}}}: For each device $k$, the energy consumption incurred by its electronic circuits during the entire service latency is computed by
\begin{equation}\label{eq:E_k_fl}
E_k^{\textrm{on}}\left(f_k,\,p_k\right) = P_k^{\textrm{on}} T_k\left(f_k,\,p_k\right)\qquad[\Joule]
\end{equation}
where  $P_k^{\textrm{on}}$ is the entire non-radiative power consumption incurred by electronic circuits those keep the device functioning. Considering practical models introduced in the existing literature, e.g., \cite{mourato18}, it will be set a realistic amount.

Thus, the total energy consumption of a device is computed by
\begin{equation}\label{eq:E_k}
E_k\!\left(f_k,\,p_k\right) = E_k^{\textrm{ex}}\left(f_k\right) + E_{k}^{\textrm{tx}}\!\left(p_k\right) + E_k^{\textrm{on}}\!\left(f_k,\,p_k\right)\qquad[\Joule]
\end{equation}

\section{Joint latency-energy optimization problem}\label{sec:Joint_Opt}
For each device $k$, the joint minimization of service latency and energy consumption is formulated by  the following bi-objective optimization problem
\begin{subequations} \label{eq:opt_T_E_k_0}
\begin{align}
& \min_{\mathbf{f},\,p_k} \;\left[T_k\left(f_k,\,p_k\right),\quad E_k\left(f_k,\,p_k\right)\right]\\
&  \quad \text{ s.t.}\;\; \sum_{k=1}^{K} f_k \le \overline{f}_0 \;\;\&\;\; 0 \le p_k \le \overline{p}_k,
\end{align}
\end{subequations}
where $T_k\left(f_k,\,p_k\right)$ and $E_k\left(f_k,\,p_k\right)$ come from \eqref{eq:T_k} and \eqref{eq:E_k}, respectively.
 As it can be seen, the devices are coupled through variables $\mathbf{f}$ in global constraint $\sum_{k=1}^{K} f_k \le \overline{f}_0$ while the other constraint is local for the device.

\subsection{Reformulation of the joint optimization problem using Tchebyshev method}
As discussed in Sec. \ref{sec:SystemModel}, the service latency and energy consumption are in general contrasting objectives. Thus, simultaneously minimizing both objectives in \eqref{eq:opt_T_E_k_0} is not possible.
Instead, one of the most widely-used methods is the Tchebyshev method that minimizes the maximum between a weighted combination of the objectives \cite{Kais99}.
To use this method, first, two parameters $\underline{T}_k$ and $\underline{E}_k$ are introduced as the minimum possible amount of energy consumption and service latency a wireless device can achieve, respectively. The amount of $\underline{T}_k$ is computed by
\begin{equation}
\underline{T}_k = \min_{p_k \le \overline{p}_k\; \& \;f_k \le \overline{f}_0} T_k\left(f_k,\,p_k\right)\qquad[\s]
\end{equation}

As service latency \eqref{eq:T_k} is a decreasing function of $f_k$ and $p_k$, it can be easily written $\underline{T}_k =  T_k(f_k = \overline{f}_0,\,p_k = \overline{p}_k)$. Similarly, $\underline{E}_k$ is computed  by
\begin{equation}\label{eq:E_k_star}
\underline{E}_k = \min_{p_k \le \overline{p}_k\; \& \;f_k \le \overline{f}_0} E_k\left(f_k,\,p_k\right)\qquad[\Joule]
\end{equation}
%
Unlike $\underline{T}_k$, it is not possible to find a closed-form solution for $\underline{E}_k$. The convexity of minimization problem \eqref{eq:E_k_star} will be discussed in the next section.

For the case of problem \eqref{eq:opt_T_E_k_0}, Tchebyshev method leads to the problem
\begin{subequations} \label{eq:opt_T_E_k_01}
\begin{align}
& \min_{\mathbf{f},\,p_k}\; \max\;\left[\eta \left(T_k\left(f_k,\,p_k\right) - \underline{T}_k\right),\quad \left(1 - \eta\right)\left(E_k\left(f_k,\,p_k\right) - \underline{E}_k\right)\right]\\
&  \quad \text{ s.t.}\;\; \sum_{k=1}^{K} f_k \le \overline{f}_0 \;\;\&\;\; 0 \le p_k \le \overline{p}_k
\end{align}
\end{subequations}
where $\eta \in (0,\,1)$ can be interpreted as the ``weight" that quantifies the desire to make each objective small or large, e.g., if one cares much less about the service latency, one can take $\eta$ small.
For a device $k$, for any $\eta \in (0,\,1)$, \eqref{eq:opt_T_E_k_01} has at least one solution that is Pareto optimal  \cite[Ch. 3]{Kais99}. Moreover, solving \eqref{eq:opt_T_E_k_01} for all $\eta \in (0,\,1)$ yields all the points on the Pareto boundary \cite{Kais99} between the service latency and  energy consumption.
On the other hand, the two extreme points
$\eta=0$ and $\eta=1$ correspond to the single-objective minimization of the energy consumption and of the service latency, respectively.
Now, let us consider its equivalent reformulation in epigraph form, namely
\begin{subequations} \label{eq:opt_T_E_k_02}
\begin{align}
&\min_{\mathbf{f},\,p_k,\,y_k} \; y_k \\
&  \quad \text{ s.t.}\;\; \sum_{k=1}^{K} f_k \le \overline{f}_0 \;\;\&\;\; 0 \le p_k \le \overline{p}_k\\
& \quad\qquad\;  T_k\left(f_k,\,p_k\right) \le \frac{y_k}{\eta} + \underline{T}_k\, ; \label{con:T_k_y_0}\\
& \quad\qquad\; E_k\left(f_k,\,p_k\right) \le \frac{y_k}{1 - \eta} + \underline{E}_k\, . \label{con:E_k_y_0}
\end{align}
\end{subequations}
where $y_k$ is an auxiliary slack variable.

\section{Convexification of optimization problem \eqref{eq:opt_T_E_k_02}}\label{sec:convexification}

\subsection{Convexification rationale}
As discussed in Sect. \ref{sec:SystemModel}, the impact of the variables $f_k$ and $p_k$ on
service latency and energy consumption is contrasting.
To find the best trade-off between the objectives, it is necessary to jointly optimize both variables for all $k$.
In the next section, we model the interaction among wireless devices using game theory, but before
applying any optimization algorithm, it is essential to ensure the convexity to reach optimality and efficiency.

Convexification is a process that transforms a non-convex optimization problem into a convex one.
While the SCA method \cite{scutari17} is a widely used technique to convexify non-convex functions,
the resulting convex approximation does not provide an accurate representation of the original non-convex function.
This is mainly due to the fact that it approximates the non-convex function with its (linear) first-order Taylor series.
In particular, the gap between the original non-convex function and its approximate convex one is
generally unknown and could be even large.
This condition is particularly critical for applications which require the optimality and uniqueness of the solution,
such as in game theory.
Therefore, our approach consists in employing a convexification technique that guarantees an equivalent
convex problem with zero gap between the optimal solutions of the original non-convex problem and its convex form
so providing unique and optimal solution.


\subsection{Convexity check and conversion}
{\boldmath\textbf{{Convexity check of $T_k\!\left(f_k,\,p_k\right)$}}}: The term $T_{k}^{\textrm{ex}}\left(f_k\right)$ is a reciprocal function of $f_k$ and thus it is a convex function of $f_k$. By neglecting inessential constant terms with respect to $p_k$, the term $T_k^{\textrm{tx}}(p_k)$, from convexity viewpoint, has the form $1/R_k(p_k)$. By exploiting the second order derivative, it is found out that $R_k(p_k)$ is a concave function in $p_k$. As $R_k(p_k)$ is a strictly positive function, its reciprocal is a convex one \cite{boyd04}. Therefore, $T_k\!\left(f_k,\,p_k\right)$ is jointly convex with respect to both variables $f_k$ and $p_k$.

{\boldmath\textbf{{Convexity check of $E_k^{\textrm{ex}}\!\left(f_k\right)$}}}: This term is a convex function since it is a square function of $f_k$.

{\boldmath\textbf{{Convexity check of $E_k^{\textrm{on}}\!\left(f_k, \, p_k\right)$}}}: This term is convex jointly on $f_k$ and $p_k$ since it is a linear function of the $T_k\!\left(f_k,\,p_k\right)$.

{\boldmath\textbf{{Convexity check of $E_k^{\textrm{tx}}\!\left(p_k\right)$}}}: By neglecting inessential constant terms with respect to $p_k$, this term has the from $1/R_k(p_k) + p_k/R_k(p_k)$. The term $1/R_k(p_k)$ is a convex function on $p_k$. While, $p_k/R_k(p_k)$ is given by the ratio between an affine and a strictly concave function with respect to $p_k$ that is neither convex nor concave \cite[Sec.~5]{avriel10}. Therefore, the function $p_k/R_k(p_k)$ has to be converted to a convex one. To do this, the following two lemmas are used: 

\begin{lemma}
Assume $\mathcal{X}$ is a compact and convex set. The solution of the minimization problem
\begin{equation}\label{eq:opt_A_B}
\min_{\mathbf{x} \in \mathcal{X}} \frac{A\left(\mathbf{x}\right)}{B\left(\mathbf{x}\right)}
\end{equation}
where $A(\mathbf{x})$ and $B(\mathbf{x})$ are positive, is the same as \cite{wang19}
\begin{equation}\label{prob:min_frac_1}
\min_{\mathbf{x} \in \mathcal{X}} \; tA^2\left(\mathbf{x}\right) + \frac{1}{4t}\frac{1}{B^2\left(\mathbf{x}\right)}
\end{equation}
where $t$ is given and it is being updated iteratively by
\begin{equation}\label{eq:update_t}
t =  \frac{1}{2A\left(\mathbf{x}\right)B\left(\mathbf{x}\right)}.
\end{equation}
\hfill $\blacksquare$
\end{lemma}

Algorithm \eqref{alg:convex_alg} summarizes the convexification of optimization problem \eqref{eq:opt_A_B}. 

\begin{lemma}\label{lemma:2}
If $A(\mathbf{x})$ is convex and $B(\mathbf{x})$ is concave, then problem in \eqref{prob:min_frac_1} is convex for given $t$ \cite{wang19}.
\hfill $\blacksquare$
\end{lemma}

 From the analysis above, instead of the non-convex term $p_k/R_k(p_k)$, it is adequate to substitute $A(\mathbf{x})$ with $p_k$ and $B(\mathbf{x})$ with $R_k(p_k)$ in Algorithm \eqref{alg:convex_alg}.
 As $R_k(p_k)$ is a concave function and $p_k$ is an affine function, given $t^q$ in each iteration $q$, the term $p_k/R_k(p_k)$ is converted to the following \emph{equivalent} convex function
\begin{equation}\label{eq:p_k_R_k_convex}
 t^qp_k^2 + \frac{1}{4t^q}\frac{1}{R_k(p_k)^2}
\end{equation}
whose convexity comes from Lemma \ref{lemma:2}.


\begin{algorithm}[t]
\LinesNumbered
\Init $q=0$; $\mathbf{x}^q = $ a feasible value; and $t^q = 1/{2A\left(\mathbf{x}^q\right)B\left(\mathbf{x}^q\right)}$\;
\Repeat{ convergence of $t^q$}{
\Compute $\mathbf{x}^{q+1} = \argmin_{\mathbf{x} \in \mathcal{X}} \; t^{q}A^2\left(\mathbf{x}\right) + \frac{1}{4t^{q}}\frac{1}{B^2\left(\mathbf{x}\right)}$\;
\Update $t^{q + 1} =  \frac{1}{2A\left(\mathbf{x}^{q+1}\right)B\left(\mathbf{x}^{q+1}\right)}$\;
\Set $q = q + 1;$ \tcp*[h]{\!\!\!\!The iteration number}\\
}
\Return $\mathbf{x}^{q};$ \tcp*[h]{\!\!\!\!The optimal solution}
\caption{Convexification of problem \eqref{eq:opt_A_B}.} \label{alg:convex_alg}
\end{algorithm}

To convexify $E_k^{\textrm{tx}}\!\left(p_k\right)$, it suffices to substitute $p_k/R_k(p_k)$ with expression \eqref{eq:p_k_R_k_convex}. The obtained equation, for a given $t$, is denoted by $E_k^{\textrm{tx}}\!\left(p_k;\,t\right)$.
Now, from the results above, the convexity of optimization problem \eqref{eq:opt_T_E_k_02} is proved.
Moreover, the total energy consumption is converted to $E_k(f_k,\,p_k;\,t)$ and finally, minimization problem of each wireless device $k$ is converted to the following equivalent convex one
\begin{subequations} \label{eq:opt_T_E_k_03}
\begin{align}
&\min_{\mathbf{f},\,p_k,\,y_k} \; y_k \\
&  \quad \text{ s.t.}\;\; \sum_{k=1}^{K} f_k \le \overline{f}_0 \;\;\&\;\; 0 \le p_k \le \overline{p}_k\\
& \quad\qquad\;  T_k\left(f_k,\,p_k\right) \le \frac{y_k}{\eta} + \underline{T}_k\, ; \label{con:T_k_y}\\
& \quad\qquad\; E_k\left(f_k,\,p_k;\, t\right) \le \frac{y_k}{1 - \eta} + \underline{E}_k \, . \label{con:E_k_y}
\end{align}
\end{subequations}

Let $\psi_k \triangleq (\mathbf{f},\,p_k, y_k)$ denotes the state of each device $k$ in convex minimization problem \eqref{eq:opt_T_E_k_03}. It also denoted by $\bm{\psi} \triangleq (\mathbf{f},\,\mathbf{p}, \mathbf{y})$ the state of all the devices where $\mathbf{y} = \{y_k\}_{k=1}^K$.
It is further denoted the feasible convex set of the state $\psi_k$ by
\begin{equation}\label{eq:S_k}
\mathcal{S}_k \triangleq \left\{\psi_k \in \reals_+^{K + 2}:\, \text{All the constraints in \eqref{eq:opt_T_E_k_03}}  \right\}.
\end{equation}
Also, the feasible convex set of $\bm{\psi}$ is denoted by $\bm{\mathcal{S}} = \prod_{k=1}^K \mathcal{S}_k$.
Minimization problem \eqref{eq:opt_T_E_k_03} is thus rewritten as:
\begin{equation}\label{eq:min_k}
\min_{\psi_k \in \mathcal{S}_k }  \; y_k
\end{equation}

 Following the weighted Tchebyshev theorem \cite[Ch. 3.4]{Kais99}, as optimization problem \eqref{eq:min_k} is convex, for every value of $\eta$ in $(0,\,1)$, it has a unique Pareto-optimal solution. Therefore, by solving optimization problem \eqref{eq:min_k} for all values $\eta$ in the range of $(0, 1)$, the Pareto boundary between the service latency and energy consumption is obtained. To model the interaction among wireless devices subject to the global constraint $\sum_{k=1}^{K} f_k \le \overline{f}_0$, cooperative game theory is used.

\section{Cooperative game formulation}\label{Sec:Coop_Game}
The fundamental difference of a cooperative approach is that, while in the non-cooperative games cooperation can only be induced as the result of matching it with self-optimization (i.e., unilateral deviations are not beneficial anyway), now the players (in this work wireless devices) are willing to cooperate, as they know that they can mutually benefit from reaching an agreement. The review of the existing literature shows that  resource allocation techniques based on cooperative games achieve much better performance and fairness compared with those based on non-cooperative games \cite{han12}.
Cooperatively binding an agreement, in game theory parlance,  is called a \emph{bargaining problem} \cite{osborne94}. The best-known bargaining solutions is NBS \cite{nash50, nash53}.
Based on NBS theorem, in a $K$-person game, if the feasible set of every player is closed, compact, and convex, optimization of the product of all players' utility functions achieves a binding solution that is unique and Pareto optimal among the objective functions $\{y_k\}_{k=1}^K$, i.e., no player can increase its payoff without unfavorably affecting the others'.
As the convexity of minimization problem \eqref{eq:min_k} on $\psi_k$ for each device $k$ is showed, NBS can be used to model the resource allocation problem.

Now, the goal can be tackled using NBS. Based on NBS theorem \cite{nash50, nash53}, cooperative utility function is defined by:\footnote{In general, the utility function of NBS is defined by $\prod_k(y_k - \underline{y}_k)$ where $\underline{y}_k$ denotes the minimum achievable $y_k$. From \eqref{eq:opt_T_E_k_01}, it is obvious that $\underline{y}_k$ is equal to zero.}
\begin{equation}\label{eq:NBS_utility}
u\left(\mathbf{y}\right) \triangleq \;\prod_{k=1}^K y_k
\end{equation}
subject to the constraints in $\bm{\mathcal{S}}$ for all $k$.
A cooperative NBS is defined as
\begin{equation}
\mathcal{G} = \left\{\left\{k\right\}_{k=1}^{K},\,\bm{\mathcal{S}},\,u\left(\mathbf{y}\right)\right\}.
\end{equation}
The cooperative NE point is computed by
\begin{equation} \label{eq:cooperative_NE_1}
\bm{\psi}^\star = \displaystyle\argmin_{\bm{\psi} \in \bm{\mathcal{S}}}
u\left(\mathbf{y}\right).
\end{equation}

In general, it is well-known that the product of convex functions does not have to be convex and thus its optimization problem belongs to a class of global optimization problems \cite{konno92}.
In non-convex global optimization, problem \eqref{eq:cooperative_NE_1} has been referred as the multiplicative optimization problem.
Whereas, it is also known that the product of positive linear functions (like the cooperative objective function $u(\mathbf{y})$), results in a quasi-convex function, i.e., it has a global optimal point \cite{benson99}. 
This leads us to relax the optimization problem \eqref{eq:cooperative_NE_1} to a problem for which known optimization techniques in the literature can be used. 

\subsection{Computing the cooperative NE}

To compute the cooperative NE of game $\mathcal{G}$, it is required to solve optimization problem \eqref{eq:cooperative_NE_1}.
To relax the optimization problem, based on the method in \cite{katoh87}, first, slack variables $\bm{\mu}= \{\mu_k\}_{k=1}^{K}$ are defined whose feasible convex set is defined as
\begin{equation}\label{eq:feasible_set_mu}
\Lambda = \left\{\bm{\mu} \in \reals^K\,|\, \prod_{k=1}^{K} \mu_k \ge 1, \quad \mu_k \ge 0\right\}.
\end{equation}
Then, the optimal solution of \eqref{eq:cooperative_NE_1} coincides with that of the following optimization problem:
\begin{equation} \label{eq:cooperative_NE}
\left\{\bm{\psi}^\star,\,\bm{\mu}^\star\right\} = \displaystyle\argmin_{\bm{\psi} \in \bm{\mathcal{S}}\;\&\; \bm{\mu} \in \Lambda}
\sum_{k=1}^{K} \mu_k y_k.
\end{equation}

\begin{algorithm}[t]
\LinesNumbered
\Init $n=0$; $\,p_k^{n} = \text{\texttt{Unif}}\left[0,\,\overline{p}_k\right]$; $\,f_k^{n} = \overline{f}_0/K$ for all $k$\;
\Compute $\underline{T}_k$ and $\underline{E}_k$ for all $k$\;
\Compute $\bm{y}^{n}=\{y_k^{n}\}_{k=1}^{K}$ using constraints \eqref{con:T_k_y} and \eqref{con:E_k_y}\;
\Repeat{ convergence of $\bm{\psi}^n$}{
\Solve convex problem \eqref{eq:cooperative_NE} given $\bm{\psi}^n$, and denote the optimal solution as $\bm{\mu}^{n+1}$\;
\Solve convex problem \eqref{eq:cooperative_NE} given $\bm{\mu}^{n+1}$, and denote the optimal solution as $\bm{\psi}^{n+1}$\;\label{line:opt_psi}
\Set $n = n + 1;$ \tcp*[h]{\!\!\!\!The iteration number}\\
}
\Return $\bm{\psi}^\star;$ \tcp*[h]{\!\!\!\!Cooperative NE}
\caption{Achieving the cooperative NE of $\mathcal{G}$.} \label{alg:iterative_alg}
\end{algorithm}

Following the NBS theorem, the existence and uniqueness of the solution of optimization problem \eqref{eq:cooperative_NE} is guaranteed since the feasible set of each minimization problem \eqref{eq:cooperative_NE_1} is convex.
Meanwhile, as it can be simply verified, the objective function of \eqref{eq:cooperative_NE} is not jointly convex on $(\bm{\psi},\,\bm{\mu})$.
In general, there is no standard method for solving the centralized and non-convex optimization problem \eqref{eq:cooperative_NE} efficiently.
As it can be perceived, it is very hard to solve the problem using standard methods due to the large space of variables.
In the following, an efficient iterative method based on block coordinate descent \cite{wright15} is proposed which minimizes the non-convex problem in \eqref{eq:cooperative_NE} by minimizing it along a subset of variables at a time. In particular, the non-convex problem is divided into two convex sub-problems using block coordinate descent technique.
By doing this,  non-convex optimization problem \eqref{eq:cooperative_NE} will be converted into two standard convex optimization sub-problems each of which can be minimized along a subset of variables at each iteration. The advantage is that
each sub-problem can be conveniently solved by standard convex optimization algorithms. Specifically, the following two standard convex sub-problems are introduced:
\begin{enumerate}
  \item For any given $\bm{\psi}$, the optimization of slack variables $\bm{\mu}$  in \eqref{eq:cooperative_NE} is a standard convex problem;
  \item For any given slack variables $\bm{\mu}$, problem \eqref{eq:cooperative_NE} is a standard convex one with respect to  $\bm{\psi}$.
\end{enumerate}
Each of two sub-problems can be minimized using standard convex optimization solutions, e.g., log-barrier method, or using existing optimization tool such as CVX \cite{cvx12} that is a Matlab-based convex optimization framework.

Algorithm \eqref{alg:iterative_alg} describes the overall iterative centralized algorithm for computing the cooperative NE of the game $\mathcal{G}$. At the initialization step $n=0$, variables $p_k$ for all $k$ take values uniformly in the interval of zero and its upper-bound $\overline{p}_k$, i.e., $p_k^{n=0} \in \text{\texttt{Unif}}\left[0,\,\overline{p}_k\right]$ where \texttt{Unif} denotes the Uniform distribution. The capacity of CPU on the fog node is equally divided among all devices, i.e., $f_{k}^{n=0}=\overline{f}_0/K$.
An iterative algorithm is proposed for the optimization problem by applying the block coordinate descent method \cite{wright15}. Specifically, the entire optimization variables in original problem are partitioned into two blocks, i.e., $\{\{\bm{\mu}\},\,\{\bm{\psi}\}\}$. Then, the variables in each block are alternately optimized, by solving standard convex optimization problem \eqref{eq:cooperative_NE}  while keeping the other block of variables fixed. Furthermore, the obtained solution in each iteration is used as the input of the next iteration. The algorithm is terminated when the fractional change of the value of the objective function is less than a threshold.  Note that the optimization of $\{\bm{\psi}\}$ in Line \eqref{line:opt_psi} is solved based on the discussion in Sec. \ref{sec:convexification} and using Algorithm \eqref{alg:convex_alg}. It is also worth pointing out that the convergence analysis for the classical coordinate descent method \cite{wright15} can be directly applied and the convergence of Algorithm \eqref{alg:iterative_alg} is guaranteed.


\section{Numerical results}\label{sec:NumRes}

Let us consider the system model described in Section \ref{sec:SystemModel}, with parameters as: $N_0 = -174$ dBm/Hz, $\overline{p}_k = 2$ W, $\overline{f}_0 = 1.2$ GHz, $\beta = -90$ dB, and $\alpha = 3.5$. To cover a wide range of IoT tasks with different sizes and different requirements for CPU capacity, we choose $D_k \sim \texttt{Unif}[0.1,\, 1.1]$ MBytes and  $C_k \sim \texttt{Unif}[50,\, 250]$ CPU-cycle/bit.
Further, based on the assessments in \cite{mourato18}, we set a practical amount to $P_k^{\textrm{on}} \sim \texttt{Unif}[2,\, 3.5]$ W.

\begin{figure}[t]
	\centering
	\psfrag{Latency}[c][c][0.9]{$T_k\left(f_k,\,p_k\right)$ [s]}
	\psfrag{Energy}[c][c][0.9]{$E_k\left(f_k,\,p_k\right)$ [Joule]}
	\psfrag{lambda27lambdal}[c][c][0.8]{\hspace{-0.93cm}$\lambda = 10^{-27}$}
	\psfrag{lambda25}[c][c][0.8]{$\,\quad\lambda = 10^{-25}$}
	\psfrag{PracticalCaseCas}[c][c][0.8]{\hspace{-0.52cm} Practical case}
	\psfrag{IdealCase}[c][c][0.78]{\hspace{1.05cm} Unrealistic case}
	\includegraphics[width=0.45\textwidth]{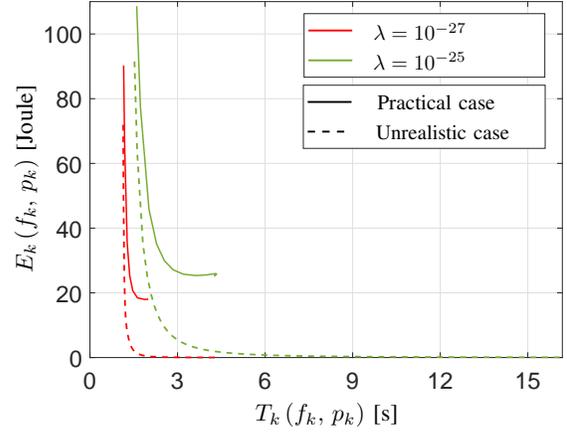}
	\caption{Pareto boundary between Energy consumption [Joule] and service latency [s] with $K=3$ and $B = 0.1 \text{ [MHz]}$  .}\label{fig:Energy_vs_Delay_K_3_B_0.1}\vspace{-0.35cm}
\end{figure}

\begin{figure}[t]
	\centering
	\psfrag{Latency}[c][c][0.9]{$T_k\left(f_k,\,p_k\right)$ [s]}
	\psfrag{Energy}[c][c][0.9]{$E_k\left(f_k,\,p_k\right)$ [Joule]}
	\psfrag{lambda27lambdal}[c][c][0.8]{\hspace{-0.93cm}$\lambda = 10^{-27}$}
	\psfrag{lambda25}[c][c][0.8]{$\,\quad\lambda = 10^{-25}$}
	\psfrag{PracticalCaseCas}[c][c][0.8]{\hspace{-0.52cm} Practical case}
	\psfrag{IdealCase}[c][c][0.78]{\hspace{1.05cm} Unrealistic case}
	\includegraphics[width=0.45\textwidth]{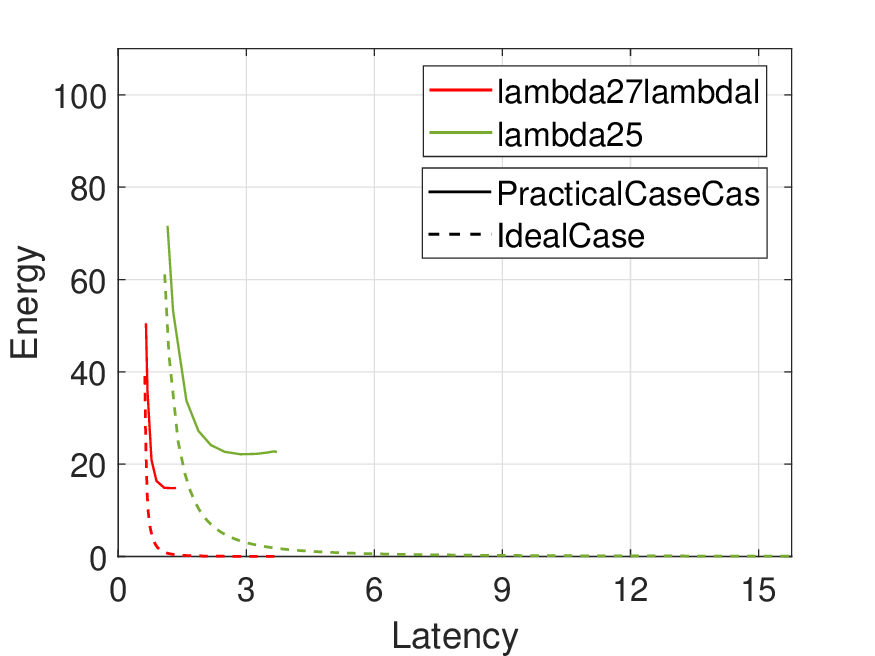}
	\caption{Pareto boundary between Energy consumption [Joule] and service latency [s] with $K=3$ and $B = 0.2 \text{ [MHz]}$.}\label{fig:Energy_vs_Delay_K_3_B_0.2}\vspace{-0.35cm}
\end{figure}

Figures \ref{fig:Energy_vs_Delay_K_3_B_0.1} and \ref{fig:Energy_vs_Delay_K_3_B_0.2} depict the Pareto boundary between energy consumption and service latency obtained with $p_k$ and $f_k$ optimized by the Algorithm \ref{alg:iterative_alg}. The results are derived from the average of $1,500$ runs in a network with $K=3$ devices randomly located in a cell with radius $70$ m. Two different settings are considered: $B = 0.1$ MHz in Fig. \ref{fig:Energy_vs_Delay_K_3_B_0.1} and $B=0.2$ MHz in Fig. \ref{fig:Energy_vs_Delay_K_3_B_0.2}.
Each figure reports four curves: two settings $\lambda = \{10^{-27},\,10^{-25}\}$ \cite{intel1, amd1} evaluated for both the practical power consumption model (solid lines) as discussed in Sec. \ref{sec:ProblemForm} and the unrealistic case where power consumption of electronic circuits is zero (dashed lines), i.e., $P_k^c = P_k^{RF} = P_k^{BB} = P_k^{\textrm{on}} = 0$, as considered in most existing works. The red lines depict the case $\lambda = 10^{-27}$ and the green lines depict $\lambda = 10^{-25}$.
As indicated by Figs. \ref{fig:Energy_vs_Delay_K_3_B_0.1} and \ref{fig:Energy_vs_Delay_K_3_B_0.2}, when the bandwidth increases in practical cases, the minimum level of the energy consumption does not change, while its impact on the maximum one is significant.
This means that increasing the bandwidth provides less energy consumption when $\eta$ is close to one.
This is consistent with the discussion in Sec. \ref{sec:Joint_Opt} that the maximum energy consumption is achieved with $\eta$ close to one.
On the other hand, increasing the bandwidth does not impact the range of service latency.
Further, the behavior of the Pareto boundary for the practical and unrealistic cases are quite different. Specifically, in the practical cases, the range of service latency is relatively short, i.e., the gap between the minimum and maximum values is limited.
It implies that, a network designer can achieve minimum energy consumption with (almost) minimum service latency  by setting $\eta$ around zero, which is an appealing configuration for the network performance.

 \begin{figure}[t]
 \centering
    \psfrag{eta09}[c][c][0.8]{\color{red}{$\eta = 0.9$}}
    \psfrag{eta01}[c][c][0.8]{\color[rgb]{0.47,0.62,0.19}{$\eta = 0.01$}}
    \psfrag{f0}[c][c][0.9]{$\overline{f}_0$ [GHz]}
    \psfrag{Latency}[c][c][0.9]{$T_k\left(f_k,\,p_k\right)$ [s]}
    \psfrag{Energy}[c][c][0.9]{ $E_k\left(f_k,\,p_k\right)$ [Joule]}
    \psfrag{ProposedSolutionSoluti}[c][c][0.8]{\hspace{-0.6cm} NBS-based solution}
    \psfrag{NoOptimization}[c][c][0.78]{\hspace{1.05cm} Equally share solution}
    \includegraphics[width=0.45\textwidth]{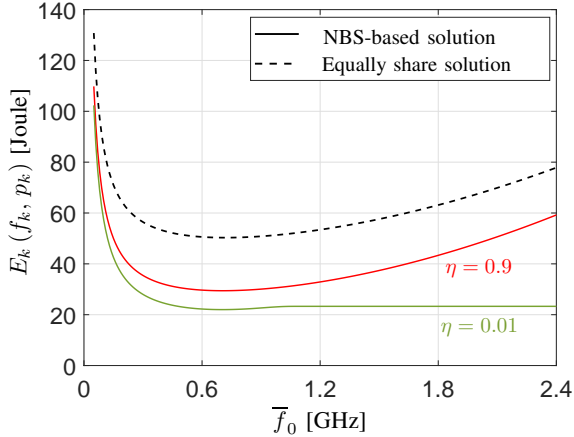}
 \caption{Energy consumption [Joule]  as a function of $\overline{f}_0$ with $\lambda=10^{-25}$, $B=0.2$ [MHz], and $K = 3$.}\label{fig:Energy_vs_Freq_K_3}\vspace{-0.35cm}
 \end{figure}

\begin{figure}[t]
	\centering
	\psfrag{eta09}[c][c][0.8]{\color{red}{$\eta = 0.9$}}
	\psfrag{eta01}[c][c][0.8]{\color[rgb]{0.47,0.62,0.19}{$\eta = 0.01$}}
	\psfrag{f0}[c][c][0.9]{$\overline{f}_0$ [GHz]}
	\psfrag{Latency}[c][c][0.9]{$T_k\left(f_k,\,p_k\right)$ [s]}
	\psfrag{Energy}[c][c][0.9]{ $E_k\left(f_k,\,p_k\right)$ [Joule]}
	\psfrag{ProposedSolutionSoluti}[c][c][0.8]{\hspace{-0.6cm} NBS-based solution}
	\psfrag{NoOptimization}[c][c][0.78]{\hspace{1.05cm} Equally share solution}
		\includegraphics[width=0.45\textwidth]{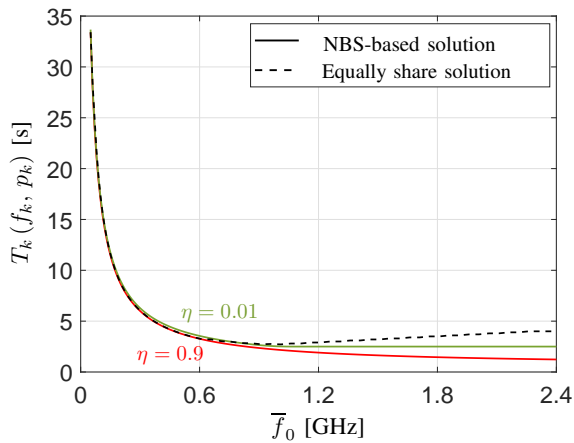}
	\caption{Service latency [s] as a function of $\overline{f}_0$ with $\lambda=10^{-25}$, $B=0.2$ [MHz], and $K = 3$.}\label{fig:Delay_vs_Freq_K_3}\vspace{-0.35cm}
\end{figure}

 Figures \ref{fig:Energy_vs_Freq_K_3} and \ref{fig:Delay_vs_Freq_K_3}  depict energy consumption and service latency, respectively, as a function of the maximum available CPU frequency $\overline{f}_0$ in a network with $K=3$, $B = 0.2$ MHz, and $\lambda = 10^{-25}$. The results of the proposed solution based on NBS (solid lines) are compared to those of a network with a classical virtualization system \cite{li18_1} (dashed lines) wherein the available CPU frequency is equally shared among wireless devices and each one transmits at the maximum power level, i.e., $f_k = \overline{f}_0/K$ and $p_k = \overline{p}_k$.
 For the proposed solution, the performance is obtained with $\eta \in \{0.01,\,0.9\}$.
 In all cases, the energy consumption and service latency achieved by the proposed NBS solution are always less than those achieved in the non-optimized network.
 To be specific, the energy consumption with $\eta = 0.01$ is less than that with $\eta = 0.9$ since the (importance) weight it has in the corresponding objective found in minimization problem \eqref{eq:opt_T_E_k_02} decreases with increasing $\eta$. This is clearly the opposite  about the service latency.
 With increasing $\overline{f}_0$, with $\eta = 0.01$ and $\eta = 0.9$, the energy consumption is achieved around $25\%$ and $75\%$, respectively, of that in a network without optimization,
 while, the achieved service latency is $60\%$ and $45\%$, respectively.
 Figures \ref{fig:Energy_vs_Freq_K_3}  and \ref{fig:Delay_vs_Freq_K_3} say that with small $\overline{f}_0$, the energy consumption is relatively high.
 By using a weak CPU, indeed, the tasks execution times increase and, consequently,  energy consumption of both wireless devices and the CPU fog node increases.
 Moreover, the energy and latency performance shown by the proposed solution are similar to those of the non-optimized network due to the lack of enough computing resource. In other words, there does not exist enough computing resource to be optimally shared among wireless devices and as a result, optimal allocation of the computing resource (almost) coincides with equally division.
 By contrast, as $\overline{f}_0$ increases, the network is able to allocate optimal CPU frequency for each task.
 As it can be shown in an optimized network, the energy consumption with $\eta=0.01$ is fixed after a certain level of maximum frequency value,
 meaning that the fog node does not necessarily use the entire CPU frequency and so a fraction of CPU capacity may remain free.

It has to be emphasized that our problem formulation, and so the results we gained, quite differs from existing literature wherein
$i)$ the bi-objective minimization approach is not investigated;
$ii)$ in most of them a fixed value is considered for the non-radiative power consumption of the electronic circuits of wireless devices, see the references cited in Sub-Section \ref{sec:Related works};
$iii)$ the latency-energy Pareto boundary is shown to be unique and obtained without any loss of optimality.
The above points justify the reasons why we have not taken into consideration any alternative solution as comparative reference.

\section{Conclusions} \label{sec:Concl}

A cross-layer solution has been proposed to allocate resources in a fog-assisted network wherein resource-constrained IoT wireless devices simultaneously offload their tasks to the serving fog node.
 The following constraints are considered: $1)$ the transmit power level of each wireless device is upper-bounded, $2)$ the computing capacity of the fog node is allocated to the wireless devices based on their requirements.

 The service latency and energy consumption being conflicting on both transmit power level of IoT devices and the CPU frequency of the fog node,
   a bi-objective minimization using Tchebyshev theorem is formulated to derive the latency-energy Pareto boundary.
  The convexity of the minimization problem of each wireless device is checked and non-convex constraints has been convexified without losing the optimality.
  The convex optimization problem for competing IoT devices is then solved exploiting
  the Nash bargaining game whose unique equilibrium point is computed by means of
    block coordinate descent algorithm.
  In order to make the problem more realistic from implementation viewpoint,
   the power consumption of electronic circuits of wireless devices is formulated using practical models.

 The results we have gained verify that:
  $i)$ the proposed resource allocation method outperforms current wireless fog networks wherein the CPU frequency is equally shared among wireless devices;
 $ii)$ using practical models for power consumption of wireless devices changes the behavior of the latency-energy Pareto boundary compared with unpractical models;
 $iii)$ increasing the bandwidth results in decreasing the range of the energy consumption interval whereas it does not impact on that of service latency.


\section*{Acknowledgement}
This work is partially supported by  POIANA (Platform for Observations In Agricultural and eNvironmentAl fields)
University of Pisa grant DR \#589 and by US NSF grant \#2128596.

\setlength{\IEEEilabelindent}{2\IEEEiedmathlabelsep}
\IEEEusemathlabelsep

\bstctlcite{mybibfile:BSTcontrol}
\bibliography{IEEEabrv,mybibfile}

\end{document}